\documentclass[prl,groupedaddress,superscriptaddress,twocolumn]{revtex4-1}

\usepackage[T1]{fontenc}
\usepackage[utf8]{inputenc}
\usepackage{lmodern}
\usepackage{graphicx}
\usepackage{amsmath}
\usepackage{amssymb}
\usepackage{amsfonts}
\usepackage{xcolor}
\usepackage[colorlinks=true, linkcolor=black, hyperindex=true]{hyperref}
\usepackage{epstopdf}

\begin{document}
\title{Atomically thin dilute magnetism in Co-doped phosphorene}

\author{L. Seixas}
  \email{seixasle@gmail.com}
	\affiliation{Centre for Advanced 2D Materials and Graphene Research Centre, National University of Singapore, Singapore 117542, Singapore}

\author{A. Carvalho}
	\affiliation{Centre for Advanced 2D Materials and Graphene Research Centre, National University of Singapore, Singapore 117542, Singapore}

\author{A. H. Castro Neto}
	\affiliation{Centre for Advanced 2D Materials and Graphene Research Centre, National University of Singapore, Singapore 117542, Singapore}
	\affiliation{Boston University, 590 Commonwealth Avenue, Boston, Massachusetts 02215, USA}

\date{\today}

\begin{abstract}
  Two-dimensional dilute magnetic semiconductors can provide fundamental insights in the very nature of magnetic orders and their manipulation through electron and hole doping. Despite the fundamental physics, due to the large charge density control capability in these materials, they can be extremely important in spintronics applications such as spin valve and spin-based transistors. In this article, we studied a two-dimensional dilute magnetic semiconductors consisting of phosphorene monolayer doped with cobalt atoms in substitutional and interstitial defects. We show that these defects can be stabilized and are electrically active. Furthermore, by including holes or electrons by a potential gate, the exchange interaction and magnetic order can be engineered, and may even induce a ferromagnetic-to-antiferromagnetic phase transition in $p$-doped phosphorene.
\end{abstract}


\maketitle

\section{Introduction}

Over the last decades the field of dilute magnetic semiconductors (DMS) has seen important developments \cite{dielt2000,ohno2000electric,ohno1998making,sato2010first,zunger2010quest,dielt2010tenyears,matsumoto2001room,dmo}, both in fundamental aspects and prospective technological applications. As for the fundamental aspects, it was possible to understand the mechanisms of interaction between diluted magnetic impurities allowing ferromagnetic semiconductors at room temperature \cite{matsumoto2001room,dmo}. From the point of view of prospective applications, the marriage between the world of two-dimensional semiconductors and the world of magnetic data storage could provide us two-dimensional spintronics devices such as non-volatile magnetoresistive memories, spin valve, spin-based transistors and even magnetically enhanced optoelectronics devices \cite{zutic2004spintronics}.

Theoretical works based upon first-principles calculations have greatly contributed to the understanding of magnetism in dilute semiconductor materials \cite{sato2010first,zunger2010quest}. Furthermore, it has been possible to achieve a spontaneous magnetic order above room temperature in a few classes of materials, including magnetic semiconducting oxides and nitrides \cite{dielt2010tenyears,dmo}. However, in practice disorder effects and thermal fluctuations result in Curie temperatures well below the expectations.

In parallel, despite the success of two-dimensional (2D) materials such as graphene \cite{novoselov2004electric,neto2009electronic}, transition metal dichalcogenides (TMD) \cite{novoselov2005two,mak2010atomically} and phosphorene, magnetism in 2D semiconductors has remained almost unexplored. One of the reasons for this might have been a theoretical result \cite{priour2005} indicating that the Curie temperature in 2D DMS tends to be substantially lower than for the corresponding 3D system. However, a subsequent work dissipated doubts by showing that including an adequate description of the disorder and the temperature dependence of the Fermi energy, it is possible to achieve much higher transition temperatures \cite{meilikhov2006}. Concordantly, a Monte Carlo study of Mn-doped MoS$_2$, with first-principles parameters, has unraveled the potential of monolayer materials by predicting Curie temperatures above room temperature \cite{ramasubramaniam2013mndoped}.

From a technical point of view, 2D semiconductors have other winning factors that can be explored in magnetic or spintronic devices. Firstly, the carrier concentration can be controlled externally by gating. And there is room for improving the control over the impurity concentration, for example exploring the possibility of using adatoms to incorporate impurities in concentrations above the solubility limit. In practice, studies in magnetic semiconductor nanostructures with lower dimensionalities, including semiconductor nanocrystals and nanowires \cite{dielt2010tenyears,zno-amines,zno-nanowire,zno-thinfilm} doped with transition metals (TM), indicate that the confinement effect and improved control of magnetic dopants can be used to increase the Curie temperature.

Phosphorene \cite{rodin2014strain,liu2014phosphorene,li2014black}, a monolayer of black phosphorus, presents some advantages compared with other previously studied 2D semiconductors. Different from semiconducting TMD, it is an elemental semiconductor and therefore less prone to vacancies \cite{liu2014two}. Further, it has an anisotropic structure allowing in principle for different exchange interaction along the two principal directions. In host heteropolar semiconductors as GaAs (zincblende structure), there is antisite defects which reduces the Curie temperatures of TM-doped GaAs dilute magnetic semiconductors \cite{sato2010first}. It is expected that in 2D heteropolar semiconductors as TMD, antisite defects \cite{zhou2013intrinsic} will also cause this effect. However, in elemental host semiconductors as phosphorene, this kind of defects does not exist. Thus, Curie temperatures of dilute magnetic semiconductors based on phosphorene are likely to be higher than those found in TM-doped TMD.

\begin{figure*}[!htb]
    \centering
        \includegraphics[width=0.96\textwidth]{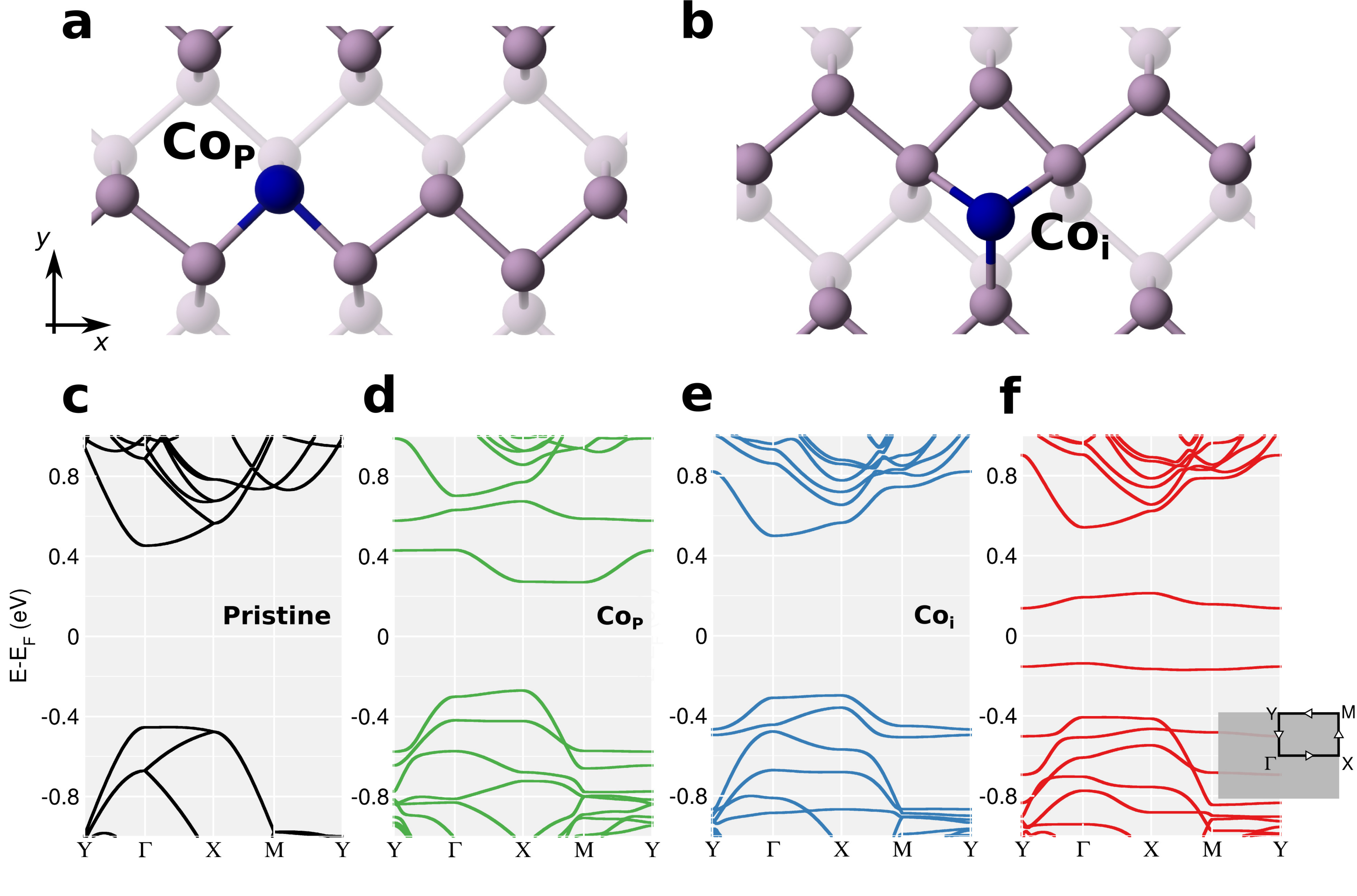}
    \caption{Ball-and-stick representation of phosphorene with: (a) substitutional Co (Co$_{\rm P}$) defect and (b) interstitial (or adsorbed) Co (Co$_{\rm i}$) defect. Electronic band structures for: (c) pristine phosphorene, (d) Co$_{\rm P}$-doped phosphorene, (e)-(f) Co$_{\rm i}$-doped phosphorene for spin up and down, respectively.}
    \label{fig:Fig1}
\end{figure*}

In this article, we use Co as a prototype magnetic impurity to show that it possible to achieve magnetism (ferromagnetism and antiferromagnetism) in phosphorene and that this can be switchable by the gate voltage through the control of the carrier density. We first studied the stability and electronic properties of Co defects in phosphorene (interstitial and substitutional), and thereafter, the magnetic orders acquired through the long-range interaction between diluted Co defects.

\section{Methods}
Electronic, energetic and magnetic properties of the systems studied were calculated via first principles methods based upon density functional theory (DFT) \cite{hohenberg1964,kohn1965}. The simulations were performed with the \textsc{Vasp} code (Vienna Ab initio Simulation Package) \cite{vasp1,vasp2} with external potentials in PAW approximation \cite{blochl1994} and exchange-correlation functional in Perdew--Burke--Ernzerhof approximation (GGA-PBE) \cite{perdew1996generalized} and Heyd--Scuseria--Ernzerhof approximation (HSE06) \cite{heyd2003hybrid}. We used cutoff energy of 400~eV for Kohn--Sham orbitals and first Brillouin zone $k$-points sampling with $5 \times 5 \times 1$ grid calculated with Monkhorst--Pack algorithm \cite{monkhorst1976} for $5 \times 4$ supercell. For Density of States (DoS) and Projected Density of States (PDoS), we use Brillouin zone integrations with $k$-points sampling in $15\times 15\times 1$ grid. The geometries were relaxed with residual force criteria of 10$^{-3}$~eV/\AA. The primitive unit cell for pristine phosphorene in our calculations was characterized by lattice constants $a \approx 3.298$~\AA\ and $b \approx 4.625$~\AA. We used vacuum distances of 15~\AA\ to avoid spurious interactions. For the calculation of magnetic anisotropy energies (MAE), we used noncollinear spin polarization in fully-relativistic framework. Volumetric data were analyzed with \textsc{Vesta} program \cite{vesta}.

Ionisation levels were calculated using the marker method \cite{markermethod}, taking the pristine supercell as reference. This method was found to give similar results to ionisation levels calculated using the formation energy method, after correcting for spurious electrostatic interactions, which partially cancel when systems with the same charge are compared \cite{carvalho-mos2}.

\begin{figure*}[!htb]
    \centering
        \includegraphics[width=0.96\textwidth]{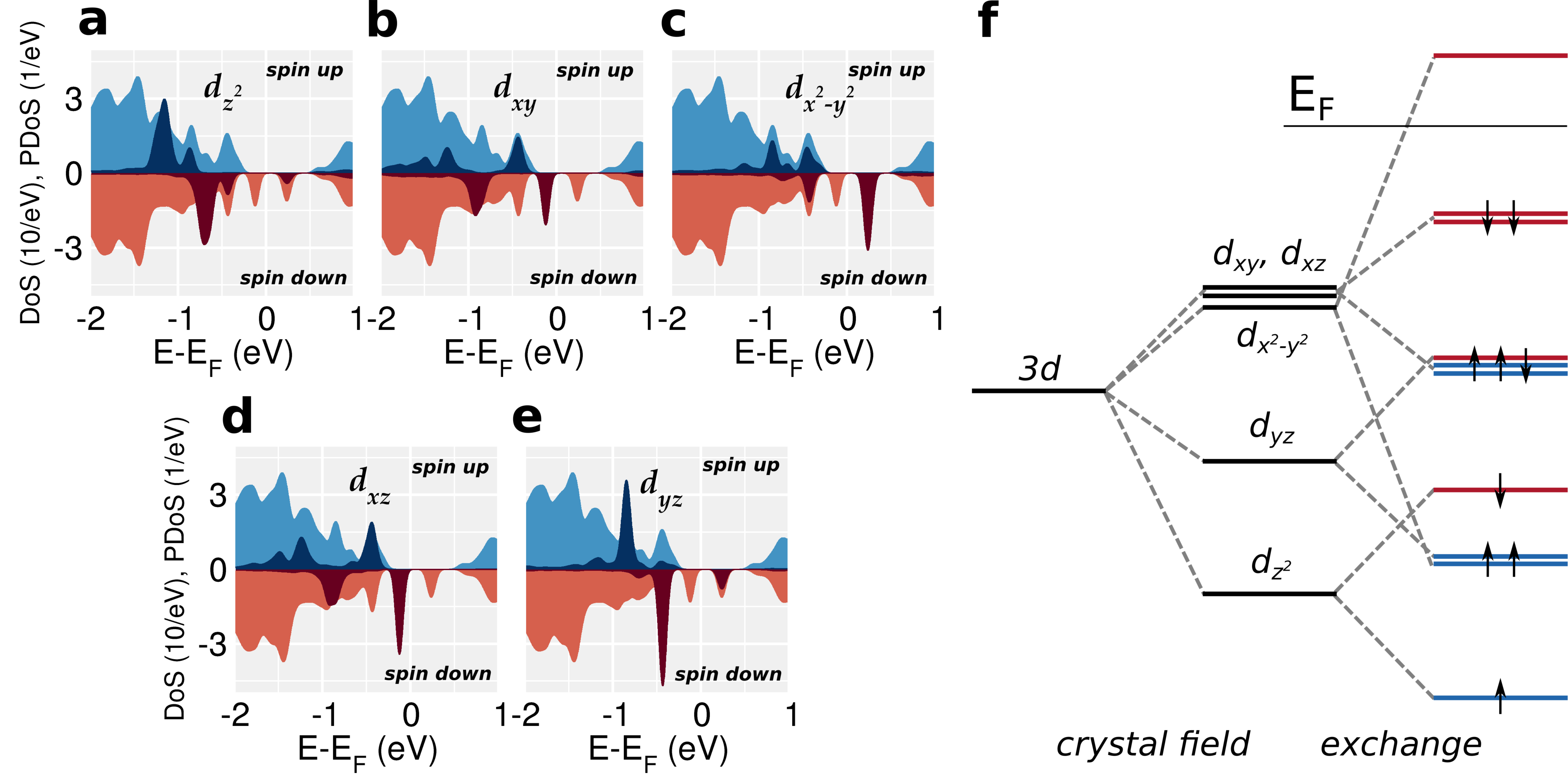}
    \caption{(color online) (a)-(e) Density of States (DoS) shown as light blue and light red areas, and Projected Density of States (PDoS) on cobalt atoms shown as dark blue and dark red areas. The PDoS are decomposed on atomic orbitals: (a) $d_{z^2}$, (b) $d_{xy}$, (c) $d_{x^2-y^2}$, (d) $d_{xz}$, and (e) $d_{yz}$. The DoS scale is ten times larger than the PDoS scale. (f) Schematic representation of $3d$ level splitting in crystal field with $C_{s}$ point symmetry and exchange interaction.}
    \label{fig:Fig2}
\end{figure*}

\section{Cobalt point defects}
Cobalt atoms can either be adsorbed on phosphorene or occupy a substitutional position (Fig.~\ref{fig:Fig1}). Both of them lead to very small lattice distortion. Substitutional cobalt ($\rm  Co_P$) forms three bonds, 3--5\% shorter than the original P-P bonds [Fig.~\ref{fig:Fig1}(a)]. Interstitial (or adsorbed) cobalt ($\rm  Co_i$) also forms three bonds, connecting P atoms in two zigzag ridges on the same side of the layer [Fig.~\ref{fig:Fig1}(b)]. The adsorption energy of Co$_{\rm i}$ defect is among the highest of TM adatoms on phosphorene \cite{kulish2015adsorption}.

In order to calculate the electronic properties of an isolated Co point defect in phosphorene, were used a $5 \times 4$ supercell, so that these Co point defects are diluted in concentrations of 1.23\% (interstitial) and 1.25\% (substitutional). Unless otherwise stated, all the calculations described in this article are in these concentrations. Whereas the Co$_{\rm P}$-doped phosphorene does not presents nonzero magnetic moments, the Co$_{\rm i}$-doped phosphorene presents a total magnetic moment of $m = 1\ \mu_{\rm B}$. This is 0.12~eV lower in energy than the nonmagnetic state (NM).

The Magnetic Anisotropy Energy (MAE) for the diluted Co$_{\rm i}$ defect was obtained from fully-relativistic and noncollinear spin-polarised calculations. The vertical spin-polarization ($z$-direction) proved to be the lowest energy, whereas spin polarizations in other directions ($x$ and $y$-direction) are related by MAE($yz$) = 1.0~meV, MAE($xz$) = 0.8~meV, and MAE($xy$) = 0.2~meV. Since the magnetic anisotropy is small, we will treat the magnetic impurities using a collinear spin-polarised density-functional formalism.

\subsection{Electronic band structures}
The band structures calculated for pristine, Co$_{\rm P}$-doped and Co$_{\rm i}$-doped phosphorene are shown in Fig.~\ref{fig:Fig1}(c)-(f). For the spin-degenerated Co$_{\rm P}$-doped phosphorene band structure [shown in Fig.~\ref{fig:Fig1}(d)], defect levels with dispersion of approximately 0.2~eV were introduced at about mid-gap, and the valence band shifts in such way that the valence band maximum (VBM) becomes located at $X$-point. This shifts in VBM will suppress the luminescence of phosphorene \cite{liu2014phosphorene,seixas2014exciton}. As for the band structures (spin up and down) of Co$_{\rm i}$-doped phosphorene, two localized levels are introduced in the spin down channel. One is occupied (below the Fermi level) and other unoccupied (above the Fermi level). For the spin up channel, all Co $d$-states are occupied, leading to an imbalance resulting in a local magnetic moment of 1~$\mu_{\rm B}$, which gives rise to a macroscopic magnetisation.

\subsection{Origin of the defect state}
The origin of the local magnetic moment in Co$_{\rm i}$-doped phosphorene can be understood by analyzing the crystal field splitting, the hybridisation between the Co valence states (4$s^2$3$d^7$) and the phosphorene host states, which have mainly $s$ and $p$ character.

From the symmetry breaking point of view, while pristine phosphorene has $C_{2v}$ point symmetry, cobalt point defects reduces the symmetry to $C_{s}$. Only a mirror plane, passing through the defect and perpendicular to phosphorene is preserved. In this lower symmetry, cobalt $d$ orbitals transform as $3A' \oplus 2A''$, where $d_{xy}$, $d_{x^2-y^2}$ and $d_{z^2}$ belong to the $A'$ representation, and $d_{xz}$ and $d_{yz}$ belong to the $A''$ representation \cite{tinkham1964}. Phosphorus $p$ orbitals splits in $2A'\oplus A''$, where $p_x$ and $p_y$ are in $A'$ representation, and $p_z$ is in $A''$ representation. We observe that the $pd$ hybridisation occurs in $A'$ with $p_x$, $p_y$, $d_{xy}$ and $d_{x^2-y^2}$ orbitals, and in $A''$ with $p_z$ and $d_{xz}$.

Different from interstitial transition metals in other materials, the hybridisation is quite strong and therefore the electronic states cannot be understood based only on a crystal field splitting model. This is apparent in the Projected Density of States (PDoS) on cobalt atoms, obtained from the decomposition on atom-centred real spherical harmonics, shown in Fig. \ref{fig:Fig2}(a)-(e). We noticed that there is $pd$ hybridisation between $p$ orbitals from phosphorene and $d$ orbitals from cobalt defects, resulting in bonding and antibonding hybrid states. The exchange field splits each of these states, leading to a splitting into four peaks for each $d$ orbital. This effect is more pronounced on $d_{xy}$, $d_{xz}$ and $d_{x^2-y^2}$ orbitals, which have maximum amplitude in the $xy$ plane [Fig.~\ref{fig:Fig2}(b),(c) and (d)]. The $d_{z^2}$ and $d_{yz}$ orbitals are nodal at the $xy$ plane and therefore undergo smaller hybridisation and have lower energies. A schematic representation of crystal field and exchange splitting is shown in Fig.~\ref{fig:Fig2}(f). For simplicity, this takes into account only the states with highest $d$ character for each $d$ orbital type and spin. Whereas the $d_{xz}$, $d_{yz}$, $d_{xy}$ and $d_{z^2}$ orbitals present exchange splitting in the range 0.32--0.45~eV, the $d_{x^2-y^2}$ orbital present exchange splitting of approximately 1.08~eV. This large apparent exchange splitting is in fact a combined effect of $pd$ hybridisation and exchange field splitting, resulting in higher peak for spin up bonding state and spin down antibonding state.

\begin{figure}[!htb]
    \centering
        \includegraphics[width=0.49\textwidth]{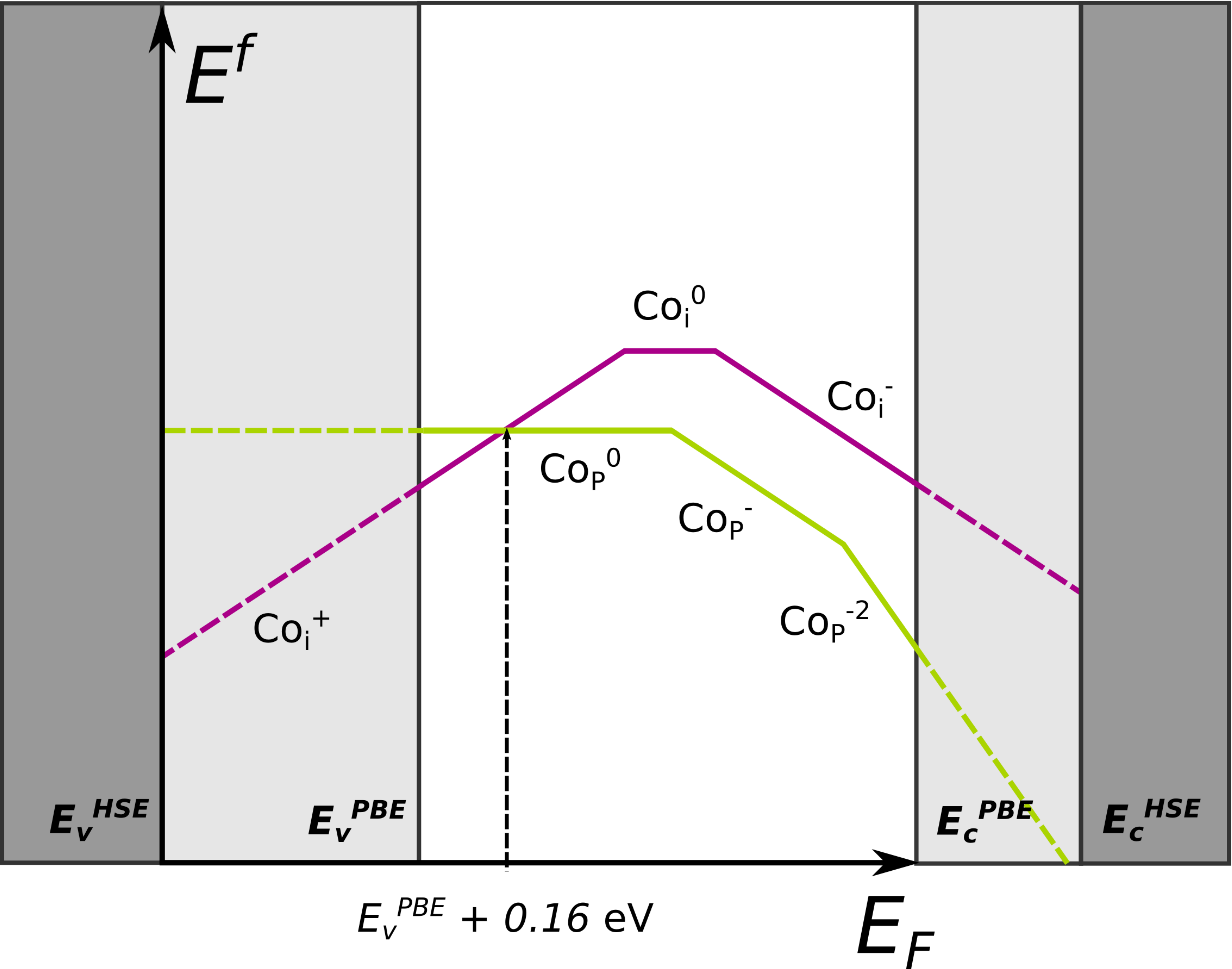}
    \caption{(color online) Formation energy vs Fermi level for charged and neutral states of Co$_{\rm P}$ (light green) and Co$_{\rm i}$ (dark violet) defects.}
    \label{fig:Fig3}
\end{figure}

\subsection{Formation energy}
We define the formation energy for neutral defects ($X^{0}$) as \cite{freysoldt2014first}
\begin{equation}
  E^{f}[X^{0}] = E_{\rm tot}[\mathrm{P:}X^{0}] - E_{\rm tot}[\mathrm{P}] - \mu_{X} - n_{\rm P} \mu_{\rm P},
\end{equation}
where $E_{\rm tot}[\mathrm{P:}X^{0}]$ is the total energy of phosphorene with $X^{0}$ defect, $E_{\rm tot}[\mathrm{P}]$ is the total energy of pristine phosphorene, $\mu_X = \mu_{\rm Co}$ is the chemical potential of cobalt calculated from 3D crystal with hexagonal lattice, $\mu_{\rm P}$ is the chemical potential of phosphorus calculated from pristine phosphorene. The number of phosphorus atoms missing ($n_{\rm P}$) is $n_{\rm P} = 0$ for Co$_{\rm i}$ defects and $n_{\rm P} = 1$ for Co$_{\rm P}$ defects.

\begin{figure*}[!htb]
    \centering
        \includegraphics[width=0.96\textwidth]{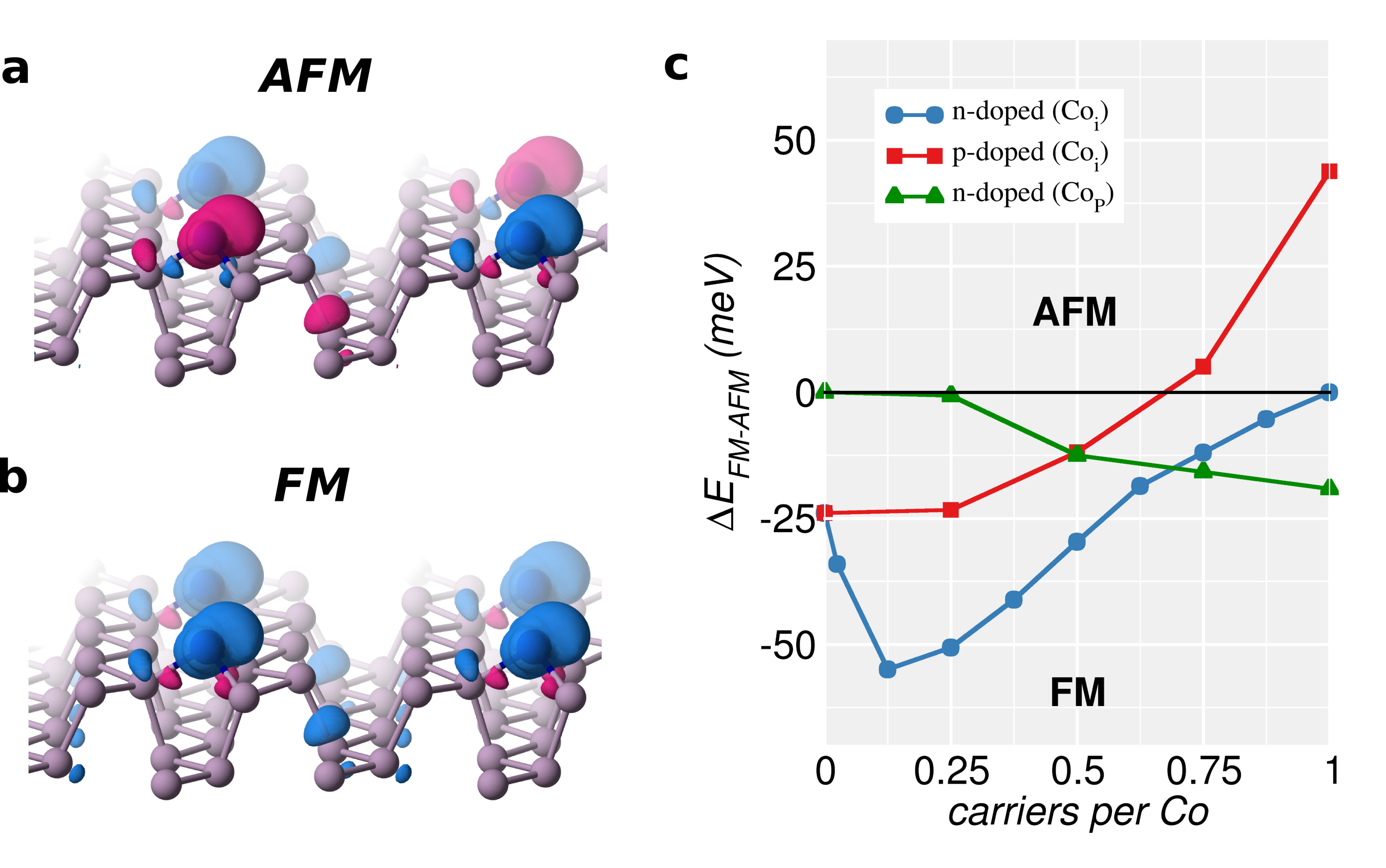}
    \caption{(color online) Spin density isosurfaces of $\pm 2\cdot 10^{-3}$ bohr$^{-3}$ for: (a) antiferromagnetic (AFM) order, and (b) ferromagnetic (FM) order. Positive values of isosurfaces are shown in blue, while negative values are shown in red. (c) Energy difference between FM and AFM orders as function of carrier densities (electrons and holes) for $n$-doped Co$_{\rm i}$ (blue), $p$-doped Co$_{\rm i}$ (red) $n$-doped Co$_{\rm P}$ (green) defects.}
    \label{fig:Fig4}
\end{figure*}

In the neutral charge state, substitutional cobalt defects have the lowest formation energy. While formation energy of neutral Co$_{\rm P}$ is $E^{f}[\mathrm{Co^{0}_{P}}] = 1.20$~eV, neutral Co$_{\rm i}$ have $E^{f}[\mathrm{Co^{0}_{i}}] = 1.42$~eV. However, both defects are electrically active and therefore the formation energy depends on the Fermi level $E_F$ (Fig.~\ref{fig:Fig3}). Calculating the thermodynamic ionisation levels $I(+/0)$ and $I(0/-)$ with the marker method \cite{markermethod}, we can evaluate the formation energies of charged point defects as a function of the Fermi level as
\begin{equation}
  E^{f}[X^{+}](E_F) = E^{f}[X^{0}] - I(+/0) + E_F,
\end{equation}
for \textit{positive} charged point defects ($X^{+}$), and 
\begin{equation} \label{eq:negative_point_defects}
  E^{f}[X^{-}](E_F) = \Delta_g^{PBE} + E^{f}[X^{0}] - I(0/-) - E_F
\end{equation}
for \textit{negative} charged point defects ($X^{-}$), where the donor and acceptor levels are given relative to the valence and conduction band edges, respectively. The Fermi level is set to zero at VBM. In equation \eqref{eq:negative_point_defects}, $\Delta_{g}^{PBE}$ is the bandgap at PBE level. Substitutional Co is an acceptor with $I(0/-)$ level at midgap (at the PBE level) and a double acceptor at $E_F = 0.78$~eV, whereas $\rm Co_i$ is amphoteric. Thus, in $p$-type material $\rm  Co_i$ is stabilised, becoming more energetically favourable than $\rm  Co_P$. According to the PBE calculations, this happens for $E_F$ between $E_v$ and $E_v+0.16$~eV. However, since the bandgap is severely underestimated at the PBE level, we extrapolated the results using the valence and conduction band shifts obtained from HSE calculations for the pristine phosphorene. Here, we assume that the gap levels of $\rm  Co_i$ and $\rm  Co_P$ are very localised, are not shifted with respect to the vacuum level by the exchange component \cite{alkauskas2008}. By fixing the PBE and HSE vacuum level at same energy, we observe the shifts $\Delta E_v = E^{HSE}_{v} - E^{PBE}_{v} = -0.48$~eV in the valence band, and $\Delta E_{c} = E^{HSE}_{c} - E^{PBE}_{c} = +0.31$~eV in the conduction band. The bandgap increase from $\Delta_{g}^{PBE} = 0.91$~eV at PBE level to $\Delta_{g}^{HSE} = 1.70$~eV at HSE level. Because of this large shift in the valence band at HSE level, the formation energy of the donor Co$_{\rm i}^{+}$ can be up to 0.42 eV smaller than the formation energy of Co$_{\rm P}^{0}$.

\section{Magnetic order}
The $pd$ hybridisation in Co$_{\rm i}$-doped phosphorene is characterized by long-range and anisotropic interactions between cobalt defects through the valence band, known as Ruderman--Kittel--Kasuya--Yosida (RKKY) interactions \cite{ruderman-kittel1954,kasuya1956,yosida1957}. These interactions have oscillating exchange coupling with the distance, leading to ferromagnetic and antiferromagnetic orders.

In order to determine the magnetic order of the fundamental state, we use a $4 \times 4$ supercell with four Co$_{\rm i}$ defects in equivalent positions. In these systems, the Co$_{\rm i}$ defects are diluted in a concentration of 5.88~\%. In neutral system, the ferromagnetic (FM) order was more energetically favourable than antiferromagnetic (AFM) order by $\Delta E_{FM-AFM} = -24$~meV. These magnetic orders are shown in Fig.~\ref{fig:Fig4}(a)-(b) by spin density ($s=\rho_{\uparrow}-\rho_{\downarrow}$) isosurfaces. Positive values of isosurfaces ($s = +2\cdot10^{-3}$ bohr$^{-3}$) are shown in blue, and negative values ($s = -2\cdot10^{-3}$ bohr$^{-3}$) are shown in red.

Since Co$_{\rm i}$ defects are amphoteric, they can accept or donate electrons. Codoping the system with electrons or holes, we can change the interaction between defects, and therefore, change the magnetic order of the system. We observe that when Co$_{\rm i}$-doped phosphorene is $n$-type, $\Delta E_{FM-AFM}$ decreases until a minimum value at 0.125$e$/Co ($n=1.7\cdot10^{13}/\mathrm{cm}^{2}$), and then increases to the ferromagnetic order is no longer stable at 1$e$/Co. For the $p$-type Co$_{\rm i}$-doped phosphorene, the difference $\Delta E_{FM-AFM}$ increases monotonically with the density of holes, so that there is a phase transition from ferromagnetic to antiferromagnetic at a critical density of holes ($p_c = 1.11\cdot 10^{14}/\mathrm{cm}^{2}$).

For similar system, but with four Co$_{\rm P}$ defects in phosphorene, due to acceptor character of Co$_{\rm P}$ defects, the magnetic orders were studied only codoping the system with electrons. In these systems, the Co$_{\rm P}$-doped phosphorene begins from nonmagnetic order at neutral state to a ferromagnetic order with $\Delta E_{FM-AFM} = -19$~meV at 1$e$/Co ($n=1.64\cdot 10^{14}/\mathrm{cm}^{2}$). Similar electric-field control of magnetism was observed in (In,Mn)As quantum well \cite{ohno2000electric}.

\section{Conclusion}
In summary, we show that interstitial and substitutional cobalt defects in phosphorene are stable, with lowest formation energy for Co$_{\rm i}$ defect in $p$-doped phosphorene, and Co$_{\rm P}$ defect in neutral and $n$-doped phosphorene. Also we show that through electron and hole codoping which can be controlled by gate potential, the exchange interaction between impurities can be engineered to yield ferromagnetic and antiferromagnetic orders. This effect can be used in prospective spintronic applications. Whereas TM-doped MoS$_{2}$ have shown diluted magnetism with very high Curie temperatures (above room temperature) \cite{ramasubramaniam2013mndoped}, dilute magnetic phosphorene presents some advantages due to absence of antisite defects and less prone to vacancies, indicating potential higher Curie temperatures and control of disordered impurities.

\section*{Acknowledgements}
L.S. acknowledges financial support provided by ``\textit{Conselho Nacional de Desenvolvimento Cient\'\i fico e Tecnol\'ogico}'' (CNPq/Brazil). The authors acknowledge the National Research Foundation, Prime Minister Office, Singapore, under its Medium Sized Centre Programme and CRP award ``\textit{Novel 2D materials with tailored properties: beyond graphene}'' (R-144-000-295-281). The first-principles calculations were carried out on the CA2DM and GRC high-performance computing facilities.

\end{document}